%Paper: hep-th/9209023
%From: A.A.Tseytlin <A.A.Tseytlin@damtp.cambridge.ac.uk>
%Date: Tue, 8 Sep 92 11:17:04 BST
%Date (revised): Mon, 21 Sep 92 19:53 BST

\input harvmac

\def \eq#1 {\eqno{(#1)}}

\def \a {\alpha}
\def \ga {\alpha}
\def \b {\beta}
\def \k {\kappa}
\def \gb {\beta}
\def \ga {\alpha}
\def \o {\omega}

\def \gp {\phi}
\def \p {\phi}
\def \ep {\epsilon}
\def \s {\sigma}

\def \r {\rho}
\def \d {\delta}
\def \l {\lambda}
\def \m {\mu}
\def \g {\gamma}
\def \n {\nu}
\def \gd {\delta}

\def \bgb {\bar \beta }
\def \gij {g_{ij}}
\def \Gmn {G_{\mu \nu}}
\def \fourth {{1\over 4}}
\def \third {{1\over 3}}
\def \e#1 {{\rm e}^{#1}}
\def \const {{\rm const }}

\def \vp {\varphi}
\def \ggij {{g_{ij}}}
\def \dg  {{\dot g}}
\def \ddg  {{ \ddot g}}
\def \df   {{\dot f}}

\def \hg {{\hat g}}
\def \B {{\hat B}}
\def \H {{\hat H}}
\def \hgg {{\hat \g}}
\def \hR {{\hat R}}
\def \tg {{\tilde g}}
\def \ha { { 1\over 2 }}
\def \dpp   {{\dot \phi }}
\def \ddp  {{\ddot \phi}}
\def \dy   {{\dot y}}
\def \hij {{H_{ij}}}
\def \dB {{\dot B}}

\def\np {  Nucl. Phys. }
\def \pl { Phys. Lett. }

\def \pr  { Phys. Rev. }

\Title{DAMTP - 92 - 49
\ \ \ hep-th/9209023}
{\vbox{\centerline{String vacuum backgrounds }
\vskip2pt
\centerline { with covariantly constant null Killing vector}
\centerline { and $2d$  quantum gravity}
}}

\centerline{
A.A. Tseytlin
\footnote{$^\dagger$}
{
 On leave of absence from the Department of
Theoretical Physics, P. N. Lebedev Physics Institute, Moscow 117924, Russia.
 \ \ e-mail: aat11@amtp.cam.ac.uk }}
\bigskip
\centerline{\it DAMTP}
\centerline{\it Cambridge University}
\centerline{\it  Cambridge CB3 9EW }
\centerline{\it United Kingdom }
\baselineskip=14pt plus 2pt minus 2pt
\vskip .3in

%abstract
We consider a  $2d$   sigma  model with a
$2+N$ - dimensional  Minkowski signature target space metric  having a
covariantly constant  null Killing vector.
We study  solutions of the conformal invariance conditions in $2+N$
dimensions  and find that generic solutions can be  represented in terms of
the RG flow in $N$ -  dimensional  ``transverse space'' theory.
The resulting conformal invariant sigma model is interpreted as a quantum
action of the $2d$  scalar (``dilaton") quantum gravity model  coupled
to a (non-conformal) `transverse' sigma model.
The conformal factor of the $2d$ metric is
identified with a light cone coordinate of the $2+N$ -
dimensional sigma model. We also discuss the case when the transverse
theory  is conformal (with or without the antisymmetric tensor background) and
reproduce  in a systematic way the solutions  with flat transverse space
known before.

\Date{9/92} %replace this line by \draft  for preliminary versions
             %or specify \draftmode at some point

%if you want double-space, use e.g. \baselineskip=20pt plus 2pt minus 2pt

\baselineskip=20pt plus 2pt minus 2pt
\newsec{Introduction}
The aim of the present  paper to give a  systematic  discussion of string tree
level vacuum backgrounds which have a covariantly constant null Killing vector.
Such (plane wave type) solutions of Einstein equations  are well known
[1] (see also [2]). Some particular examples of such spaces were found
to be  solutions of the string effective equations to all orders of
perturbation
theory in $\a'$ [3--8]. In a recent paper [9] we have considered the most
general
metric with a covariantly constant null Killing vector and  have shown that if
the ``transverse" part of the metric satisfies a first order renormalisation
group - type equation there exists a dilaton field such that the
metric--dilaton
background solves the string equations to all orders in $\a'$.
Below we shall complete the proof given in [9] and  explain  the
relation to particular solutions of
 refs. [3--8].

Part of our interest in Minkowski signature  string backgrounds with a null
Killing vector is motivated by the observation  that the corresponding sigma
model can be interpreted as describing  a model of  $2d$ quantum gravity  with
an extra scalar  field  coupled to  $2d$
curvature (see e.g. [10--13,9]).  As we shall see below,
 our solutions provide examples of consistent
quantisation of $2d$ scalar quantum  gravity  coupled to a non-conformal matter
being exact ($2+N$ dimensional) conformal theories
satisfying proper ``initial  conditions".
This is to be contrasted to the case of pure $2d$ gravity (without an
additional
scalar field) where similar description
in terms of an $1+N$ dimensional conformal theory is not known explicitly.

We shall start  in Sec.2 with  a discussion of the general form of the
``null" metric  and remaining freedom of coordinate transformations [1]. Then
we
shall study the structure  of the   sigma model Weyl invariance conditions for
the backgrounds with null Killing vector.  Using the fact
that the Weyl invariance coefficients
(``$\bar \b $ - functions") satisfy certain
identities [14] reflecting the freedom of coordinate transformations in the
target space (and related to the fact that ${\bar \b }$'s can be derived from a
covariant effective action [15,16]) we shall prove that the resulting
metric - dilaton  equations always have a solution.

In Sec.3 we shall consider the case when the ``transverse"
theory is Weyl invariant (in particular, when the ``transverse"  metric is flat
[3--8]).  We shall discuss  a number of explicit solutions, including ones with
a
non-vanishing  antisymmetric tensor.

A relation to $2d$ quantum gravity  models will be  discussed in Sec.4.

\newsec { Structure of the Weyl invariance conditions and  generic  solution}
1.  Let us consider the $D=N+2$ dimensional space with Minkowski signature. The
most general metric admitting a covariantly constant null Killing
vector\foot{We shall refer to such metrics  as ``null metrics".} can
be represented in the form  $$ds^2 = \hg_{\mu \nu} dx^{\mu} dx^{\nu} =  -2dudv
+  \ggij (u,x) dx^i dx^j   \ \  ,
 \ \eq{2.1} $$ $$ \ \ \mu , \nu = 0,1, ..., N, N+1 \ \ , \ \ \
  i,j = 1,...,N \ \ . $$
In fact, starting from the  null metric [1]
$$ds^2 = \tg_{\mu \nu} dx^{\mu} dx^{\nu} =  -2dudv +  \ggij (u,x) dx^i dx^j
 + 2A_i(u,x) dx^i du + K(u,x) du^2  \
\  ,
 \ \eq{2.2} $$
one can eliminate $A_i$ and $K$ by a change of coordinates which preserves the
``null" structure of (2.2) [1].  If
$$ x^i = f^i (u,x')\ \ , \ \ \ v=v' + h(u,x') \
\ , \eq{2.3} $$
we get
 $$ \ \  A_m' = f^i,_{m} A_i + \ggij f^i,_{m }\df^j
 - h,_{m} \ \ \ , \ \ \  K' = K + \ggij \df^i\df^j - 2\dot h  \ \ , \eq{2.4} $$
$$  g_{mn}' = f^i,_m f^j,_n g_{ij} \ \ ,
\ \ \ f^i,_m = { \del f^i \over \del {x^m}' }\ \ \ , \ \  \dot f =
{ \del f\over \del u }  \ \ . $$
It is clear that redefining $v$ one can always  absorb $K$ into a
longitudinal
part of $A_i$ (or vice versa). The equations $A_m'=0\ , \ K'=0$ are first order
differential equations  in $\del \over \del u$   for $f^i$ and $h$ and  so they
always have a solution (assuming one chooses the initial values in such a way
that the matrix $f^i,_n$ is non-degenerate) [1].

Thus the most general null metric is parametrized by the functions
$\ggij (u,x)$. It
is important to keep in mind, however, that (2.1)  considered as a generic
form of the metric is written using a special choice of coordinates $v, x^i$.
For example, if $\ggij(u,x)$ is a flat metric as a function of $x^i$ this does
not imply that  a generic  ``null" metric with a flat transverse part is just
given by $ ds^2 =-2dudv + dx^idx_i \ $: transforming  the coordinates to
make $\ggij$  equal to  $\gd_{ij}$  we will get back   the metric (2.2) with
non-vanishing  $A_i$ and $K$.

The metric  (2.1) is a natural starting point for
a discussion of general properties of solutions while  (2.2) may be  used if
one looks for a solution with a specific ansatz for $\ggij (u,x)$
(e.g. a standard metric of a flat space or a  sphere, etc).  All solutions can
of
course be expressed in any of the two equivalent forms (2.1) or (2.2).

In [9] we have studied  string  vacuum backgrounds represented by the metric
(2.1). To  reproduce the  simplest exact solutions considered in
[3--8] (which correspond to the ``null" metric with  flat transverse part) it
will be useful to make a coordinate transformation  to put the metric into the
``non-diagonal" form  (2.2).

2. Let us now discuss  the  structure of equations which we are going to
solve using the ansatz (2.1). We shall follow the notation of refs.[9,16].
The conditions of Weyl invariance of a string sigma model (parametrized by a
metric $\Gmn$ and a dilaton $\p$) are equivalent to the tree level
string   effective equations  and have the following  general form [15,17,18]
$$ {{\bar \gb}^G}_{\mu \nu } =
\gb^G_{\mu \nu} +   D_{( \mu} W_{\nu )} + 2\ga'  D_{\mu} D_{\nu} \phi =0 \ \ ,
\ \ \eqno {(2.5)} $$
$$ \bgb^{\phi} = \gb^\phi + \ha W^{\mu} \del_{\mu}\phi  + \ga' (\del_\mu
\phi)^2
  =0  \ \ , \ \eqno {(2.6)} $$
 $$ \gb^G_{\mu \nu}= \a' R_{\mu \nu} +  O(\a'^2)
  \ \ , \ \ W_\mu = O(\a'^3) \ \ , \eqno {(2.7)}$$
$$ \gb^\phi = c  -\g \phi + \o \ \ ,  \eqno (2.8) $$ $$ \ \ \g=
\sum_{n=2}^{\infty}
 M^{\m_1 ... \m_n} D_{\m_1} ...D_{\m_n} = \ha \a'D^2 +
O(\ga'^3) \ \ ,  $$
$$ \ \ \o = O(\a'^2) \ \ ,  \ \ \ c={1\over 6 } ({ D} -26) \
\ .  $$
  $\gb^G_{\mu \nu}$, $\ \g$, $\ \o$ and $W_\mu$  are covariant
functions constructed from  the curvature and covariant derivatives.
Let us note that there exists
a renormalisation scheme (i.e. a  definition of $\Gmn$) in which the
leading  order term in the `anomalous dimension' differential operator $\g$
is given just by its leading order term $\ha \a'D^2 $ [17]. The Weyl anomaly
coefficients  ${{\bar \gb}^G}_{\mu \nu }$  and $\bgb^{\phi}$ satisfy  $D$
differential identities which can be derived from the condition of
non-renormalisation of the trace of the energy-momentum tensor of the sigma
model [14].  They can be interpreted as being  a consequence of the target
space
reparametrisation invariance given that
 ${{\bar \gb}^G}_{\mu \nu }$  and $\bgb^{\phi}$  can be derived from a
covariant effective action $S$ [15,17]
$$ {\d S \over \d \vp^A} = k_{AB} \bgb^B \ \ , \ \ \vp^A=(\Gmn\ , \ \p)\ \
,\eq{2.9} $$
$$ 2D_\mu {\d S \over \d \Gmn} - {\d S \over \d \p } D^\nu \p =0 \
\ . \eq{2.10} $$
To the lowest order in $\a'$ one finds [15,17]
$$ \del_\mu \bgb^{\phi} - {{\bar \gb}^G}_{\mu \nu } D^\nu \p + \ha D^\nu (
{{\bar \gb}^G}_{\mu \nu } - \ha \Gmn G^{\l \r}{{\bar \gb}^G}_{\lambda \r}
)  + O(\a'^2) =0 \ \ . \eq{2.11} $$
In general, the identity  has the following structure [14,17]
$$ \del_\mu \bgb^{\phi} - {{\bar \gb}^G}_{\mu \nu }D^\nu \p
  - V_\mu^{\a \b}{{\bar \gb}^G}_{\a \b  }  =0  \ \ , \eq{2.12} $$
where the differential operator $V_\mu^{\a \b}  $ depends only on $\Gmn$.

One of the consequences of (2.12) is that $\bgb^{\phi}=\const$  once
(2.5) is satisfied. In general,  the identity (2.12)  implies  that  only $ \ha
D
(D+1) +1 - D$ of equations (2.5), (2.6) are independent.  It may happen, in
particular, that  if  the ``transverse" subset of $\ha (D-2)(D-1) $  equations
in (2.5) and  the dilaton equation (2.6) are solved, the remaining $D$
equations
(2.5) are satisfied automatically. This observation will be important below.

3. Let us now look for solutions of (2.5),(2.6) which have the form [9]
$$ \Gmn = \hg_{\mu \nu} (u,x) \ \ , \ \ \ \p = \p (v,u,x) \ \ , \ \ x^\mu=
(v, u, x^i) \ \ , \eq{2.13} $$
where $\hg_{\mu \nu}$ is given by  (2.1).  The non-vanishing
components of the Christoffel connection and the  curvature of $\hg$  are
  $$ {\hat \Gamma}^i_{jk} = { \Gamma}^i_{jk}\ \ ,
\ \ {\hat \Gamma}^v_{ij}=\ha \dg_{ij}
\ \ , \ \ {\hat \Gamma}^i_{ju}=\ha g^{ik}\dg_{kj}   \ \ , \ \
\ \dg_{ij}\equiv{\del \ggij \over \del u}\ \ , \  \eqno (2.14)  $$
 $$ {\hat R}_{ijkl} = { R}_{ijkl}\ \ , \ \
{\hat R}_{iuju}= T_{ij} \ \ , \ \ {\hat R}_{uijk} =  E_{ijk} \ \ , \eq{2.15} $$
 $$ T_{ij}\equiv -{1\over 2}  (\ddg_{ij}-\ha  g^{mn}\dg_{im}\dg_{nj}) \ \ , \
\ E_{ijk}= - D_{[j}\dg_{k]i}  \ \ . \eqno (2.16) $$
The `covariant' form of (2.14)--(2.16) is
$$
{\hat \Gamma}^\l_{\mu \nu} = { \Gamma}^\l_{\mu \nu } +  g^{\l \r} \dg_{\r (\m}
l_{\n)} - \ha \dg_{\m \n } l^\l  \ \ , \eq{2.17} $$
$$ {\hat R}_{\m \n \r \s} =  R_{\m \n \r \s} + 2 l_{[\m } E_{\n] \r \s }
+ 2 l_{[\r } E_{\s] \m \n } - 4 l_{[\m } T_{\n] [\r} l_{ \s ] }
\ \ , \eq{2.18} $$
$$ {\hat R}_{\m \n } =  R_{\m \n } - 2 g^{\r \s} E_{\r \s ( \m} l_{\n )}
+ l_\m l_\n g^{\r \s} T_{\r \s} \ \ ,  \eq{2.19} $$
where $l_\m = \del_\m u = (0,1,0,...,0) $ is  the null Killing vector and
$g_{\m \n}\
, \ { \Gamma}^\l_{\mu \nu }\ , \  R_{\m \n \r \s}\ , \  T_{\m \n } $ and $
E_{\m \n \l }$  have only transverse $(i,j,...)$ components being
non-vanishing.

Since $\gb^G_{\mu \nu} \ , W_\m $  and hence
 $${\gb^G_{\mu \nu}}'\equiv \gb^G_{\mu \nu} +   D_{( \mu} W_{\nu )} \ \
 $$  in (2.5) are  covariant functions of the curvature and its
derivatives we have   $$l^\m{\gb^G_{\mu \nu}}'  \equiv 0\ \ \  , \ \ \  l^\m
W_\m  \equiv 0 \ \ ,  $$ i.e. the  $( \m v)$  components of (2.19) are
identically zero. Then (2.5) gives the following constraint on the dilaton   $$
\del_\m \del_v \p =0  \ \ , \ $$ i.e.
$$ \p = p v +  \p (u,x) \ \ , \   \ \ \  \ p=\const  \ \ . \eq{2.20} $$
Here $p$ is an arbitrary integration constant
 and $\p (u,x)$ is to be determined. From now on all the functions  will depend
only on $u$ and $x^i$. Using (2.14), (2.20)  we can represent the non-trivial
components of (2.5)  as follows ($\a'=1$)
$$ \bgb^g_{ij}  - p \dg_{ij} =0 \ \
,  \ \ \eqno {(2.21)} $$
$$
 \bgb^g_{ij} \equiv \gb^G_{ij} +   D_{(i} W_{j )} + 2  D_{i} D_{j} \phi
\ \ ,  \eq{2.22} $$
$$ \gb^G_{iu} + \ha \del_i W_u + \ha \dot W_i - \dg_{ij}W^j + 2 \del_i\dpp
-\dg_{ij} D^j \p =0 \ \ ,  \eq{2.23} $$
$$ \gb^G_{uu} + \dot W_u  + 2 \ddp
=0 \ \ .  \eq{2.24} $$
Equation (2.6)  takes the form
  $$ \bgb^{\phi} = {1\over 3 } + {\bgb^{\phi}}{}' + \ha  p M^{ij} \dg_{ij}
 - \ha p W_u - 2p \dpp
 =0  \ \ , \ \eqno {(2.25)} $$
$$ {\bgb^{\phi}}{}' \equiv  c'  - \g' \p + (\del_i \p)^2  + \ha W^i \del_i
\p  +\o \ \ , \ \ c'={ 1\over 6}  (N-26) \ \ . \eq{2.26} $$
Note that being  scalar functions of the curvature (2.15),(2.18) $\g'$,  $\o$
and
hence $ \bgb^{\phi}{}'$ do not depend on the  derivatives of the metric over
$u$. The
term  $\g \p $ in (2.8)   reduces to $\g' \p - \ha p M^{ij} \dg_{ij}$  where
$\g'$ corresponds to the ``transverse" theory (i.e. contains only derivatives
over $x^i$) and the correction is due to $\del_v \p$ ($M^{ij} = \ha g^{ij} +
...$).

  The functions $\bgb^g_{ij}$
(2.22) and ${\bgb^{\phi}}{}'$ (2.26)  can be interpreted as the Weyl anomaly
coefficients of the ``transverse" theory defined by $g_{ij}(u,x)$ and $\p
(u,x)$
at fixed $u$ (1/3 in (2.25)  corresponds to the  central charge  contribution
of
the two light-cone  coordinates).

 4.  Let us first consider the case of $non-vanishing$  $p$.
Then (2.21) is a first order differential  equation for
$g_{ij}(u,x)$ which always has a solution. Eliminating the $u$ - derivatives of
$g_{ij}$ from (2.25) using (2.21) we  find  a similar first order  equation
for $\p (u,x)$. Eqs. (2.21),(2.25) can be interpreted as renormalisation group
equations of the ``transverse" theory with $u$ playing the role of the RG
``time" [9].

  The remaining question is whether the solutions of (2.21) and (2.25)
satisfy also (2.23) and (2.24). It  can be answered  positively using the
identity  (2.12).  Substituting  $ \bgb^G_{ij}=0\ , \ \  \bgb^\phi =0$ and the
expression  (2.20) for the dilaton into (2.12) we get $$  p \bgb^G_{i u}  - 2
V_i^{ju}{{\bar \gb}^G}_{j u  } - V_i^{uu}{{\bar \gb}^G}_{u u}=0  \ \ \ ,
\eq{2.27} $$ $$  p \bgb^G_{u u} - \bgb^G_{u i} D^i \p - 2V_u^{ju}{{\bar
\gb}^G}_{j u}  - V_u^{uu}{{\bar \gb}^G}_{u u  }=0\ \ . \eq{2.28} $$ Given that
$
V_\l^{\m \n }$ is a differential operator constructed from the curvature
 (2.18), its components $V_i^{ju}\ , \ V_i^{uu}$ and $V_u^{uu}$ should vanish
identically ($V_u^{ju}$ may be non-vanishing because of possible  $ V^\m
\d_\l^{\n }$ term in $ V_\l^{\m \n }$). As a result, equations (2.27),(2.28)
take
the form $$  p \bgb^G_{i u} =0\ \ , \ \ \ \eq{2.29} $$ $$
 p \bgb^G_{u u} - \bgb^G_{ iu} D^i \p - 2 V_u^{ju}{{\bar \gb}^G}_{ju}
=0\ \ . \eq{2.30} $$
 In the leading order approximation (2.11)  the identity (2.30) is given by
$$  p \bgb^G_{u u} - \bgb^G_{ iu} D^i \p + \ha D^i \bgb^G_{iu}
=0\ \ . \eq{2.31} $$
We conclude that once (2.21) and (2.25) are satisfied
for a non-zero $p$ (2.29)
and (2.30) imply that
$$ \bgb^G_{i u}=0 \ \ , \ \ \ \ \bgb^G_{u u} =0  \ \  . $$
 What we have  found  is that given  some  initial data $( \gij (x)\ , \ \p (x)
)$ at $u=0$ there exists  a  $u$ - dependent solution $ ( \gij (u,x)\ , \ \p
(u,x) ) $ of the Weyl invariance conditions (2.5),(2.6).  If the initial
transverse theory  is  generic, i.e. $\bgb^g_{ij}$ in  (2.22)  is non-vanishing
at $u=0$ then  the  solution exists only for a non-zero $p$.   If, however, the
initial theory is Weyl invariant, i.e.
 $$\bgb^g_{ij}(u=0) =0 \ \ \ , \ \  \ \ {\bgb^\p }{}' (u=0) =
c'' = \const \ \ , \ \eq{2.32} $$
  we have an option.    For $p\not=0$   the
simplest solution   of  (2.5),(2.6) is the `direct
product'  one    represented by  the fixed point
 of the RG equations (2.21),(2.25)
  $$ \gij (u,x)=\gij (x) \ \ ,
  \ \ \ \p (u,x) = {1\over 2p}( {1\over 3 } + c'')
  u + \p (x)  \ \ . \eq{2.33} $$
There  may  be   also  more interesting solutions corresponding to
interpolations
between different conformal points (one theory at $u=-\infty$ and another -- at
$u= + \infty $).  The case of $p=0$  will be discussed in the next section.

Let us note that
 if one  looks   for  special solutions with $\gij (u,x)$  corresponding to
a
conformal theory {\it for all} $ \ u$ then  it is necessary to  set $p=0$.
In fact, let us  try to find, for example, a
solution  of (2.21)--(2.26) with the metric  $\gij (u,x)$ being flat  for
arbitrary $u$ assuming $p \not= 0$. Since the transverse components of the
curvature (but not necessarily of the connection) are zero,    $\gb^G_{ij}$ and
$ W_{\m }$   vanish\foot{It is easy to see that $W_u$ must be zero  since as
follows from (2.18) the only non-vanishing contributions could come from the
terms which are linear in curvature but such terms are absent in $W_u$ [17].}
and  so   the equations for $\gij$ and $\p$ (2.21) and (2.25) take the form
 $$ p
\dg_{ij} =  2  D_{i} D_{j} \phi \ \ , \eq{2.34} $$
$$ { c} - \ha D^2_i \p + (\del_i \p)^2  + \fourth  p
g^{ij} \dg_{ij}    - 2p \dpp
 =0  \ \ , \ \ \ c= { 1\over 6}  (N-24)\ \ .   \eqno {(2.35)} $$
Eliminating $\dg_{ij}$ from (2.35) we get
$$ 2p \dpp = g^{ij} \del_i \del_j \p  + c \ \ \ . \eq{2.36} $$
Since $p\not=0$  the remaining equations (2.23), (2.24) are  again satisfied as
a
consequence of (2.34),(2.36).   Equations  (2.34),(2.36)  must be supplemented
by the condition (${ R }_{ijkl}=0$) that $\gij$ remains flat for all $u$. Since
the $u$ - derivatives of ${ R }_{ijkl}=0$ vanish automatically as a consequence
of (2.34) (note that ${\dot \Gamma}^i_{jk} = p^{-1} D^iD_jD_k \p $, etc)
the flatness condition  is only a constraint on the initial data $g_{ij}
(0,x)$ for (2.34),(2.35).
   The
resulting system, however, has only the trivial solution: it is
straightforward to check that all the components of the curvature (2.18) vanish
($E_{ijk}=0\ , \ T_{ij}=0)$. The  corresponding  metric (2.1) is flat so
that the solution for the dilaton must be linear in proper coordinates. In
fact,
$\ddot \p=0$  and the $u$ dependence in
$\gij$ can be represented  in terms of  a coordinate transformation.

To get a non-trivial solution  with a  flat $\gij (u,x)$ (or, more generally,
conformal transverse theory) one should set $p=0$.  Then  (assuming $\p = \p
(u)
$) eqs.(2.21),(2.25)   are  satisfied automatically  but since $p=0$ the
identities (2.29),(2.30)  {\it no longer imply}  that  (2.23),(2.24) are also
satisfied. To  make the analysis  of the solutions of  the  remaining equations
(2.23),(2.24)  more transparent  it is useful to change  coordinates, trading
the functions $\gij (u,x)$ corresponding to a flat transverse metric  for $A_i$
and $K$ in (2.2) (i.e.   transforming the metric (2.1) into
 (2.2) where $\gij (u,x)$ has canonical $\gd_{ij}$ form).  This will be
discussed in the next section.

\newsec { Solutions  with  conformal ``transverse" part}
1. Let us  return to the discussion of  the case  (2.32) when
the ``transverse" theory $ (\gij (u,x) \ ,  \ \p (u,x)) $ is Weyl invariant at
$u=0$,  assuming now that $p=0$.
  If we are looking for a solution of the  Weyl
invariance conditions for a  $N+2$ - dimensional background  (2.1) then for
$p=0$  eqs.(2.21),(2.25)  imply that  the  initial Weyl invariance conditions
(2.32) are satisfied  also  for  all  other values of $u$. Therefore  a
solution with  (2.32),(2.20) and $p=0$  may exist only if the transverse theory
is conformal for all $u$.
 Since  (2.21) holds
identically  it no longer gives an equation for $\gij (u,x)$. The same is
true for (2.25): it does not contain  terms with $u$ - derivatives  and being
a constant (as a consequence of (2.5),(2.12)) it is satisfied  for  all $u$
if it  is   for $u=0$, i.e. if $ \third + c' =0$.
As we already mentioned, for $p=0$   the identities
(2.12) or  (2.27),(2.28)  do not imply that eqs.(2.23),(2.24)  are  satisfied
automatically. Instead of $N+1$ identities  (2.29),(2.30)    we are left with
just one (2.30). As a result,  we get $N$ equations (2.23),(2.24) ((2.30) gives
a relation  between   components of (2.23)) for $\ha N(N+1) + 1 $ functions
$\gij (u,x)\ , \ \p (u,x)$.

Using (2.7),(2.14)--(2.18) we can represent the leading terms in (2.22),(2.23)
in
the form [9]
$$ g^{jk} E_{jik} + \ha \a' E_{mnk} R_i^{mnk}
+ 2 \del_i\dpp -\dg_{ij} D^j \p  +O(\a') $$
$$ = \ha ( D^j \dg_{ij} - \del_i(g^{jk } \dg_{jk}))  + 2 \del_i\dpp
-\dg_{ij} D^j \p  +O(\a') =0 \ \ ,  \eq{3.1} $$
$$  g^{ij}T_{ij} + 2 \ddp + O(\a') = - \ha
( g^{ij} \ddg_{ij}-\ha  g^{ij} g^{mn}\dg_{im}\dg_{nj})+ 2 \ddp + O(\a')
=0 \ \ .  \eq{3.2} $$
The count of powers of $l_\m$ in (2.17),(2.18) implies that higher order terms
in (3.1) will  be proportional to  the $D_i$ - derivatives  of  the first power
of  $\dg_{ij}$ (originating from the $O(l_\m)$ terms in the connection  (2.17)
or  the curvature (2.18))  multiplied by factors  constructed  from  $D_i$ and
$R_{ijkl}$.  In a similar fashion, the  higher order terms in (3.2) will be
linear in   $D_i$ - derivatives of $\ddg_{ij}$ or  quadratic in $D^s
\dg_{ij}$.

It is easy to check that the identity (2.31) is indeed satisfied by the leading
terms in (3.1). Solving  formally (3.2) for the dilaton and  substituting the
result into  (3.1) one  finds  a system of $N$  equations  for $\gij (u,x)$
with one identity (2.30).

2. Let us  now make a specific assumption. Since  in any case only $N-1$   of
$\ha N(N+1)$ components of  $\gij $ are constrained by (3.1),(3.2)  let us
assume that the solution $\gij (u,x)$
 can be represented as a  $u$ - dependent  coordinate ``rotation"
of  a $u$-independent (e.g. flat)  metric
$$ \gij (u,x) = \del_i y^m \del_j y^n g_{mn} (y) \ \ , \eq{3.3} $$ $$ \ y^m =
y^m (u,x)\ \ , \ \ y^m(0,x)=x^m\ \ , \ \ \gij (0,x) = \gij (x) \ \ .  $$
Substituting (3.3) into  (3.1),(3.2) we get  a  system of equations $y^i(u,x)$
and $\p (u,x)$.  In order to simplify the subsequent analysis  let us
 first ``undo" the coordinate transformation in (3.3).
Represented in terms of the new coordinates $y^i$ the metric (2.1)  takes the
non-diagonal form (2.2)  with  the $u$-independent transverse metric (cf.(2.4))
$$ \gij (u,y)= \gij (y)
\ \ , \ \ \  A_i(u,y) = - \gij \dy^j \ \ , \ \  \  K(u,y) = \gij \dy^i \dy^j \
\ . \eq{3.4} $$
Instead of solving (3.1),(3.2) for $y^i(u,x)$
we shall  consider  eqs.(2.5) for the metric (2.2),(3.4)  $G_{\m \n } = \tg_{\m
\n } $ and solve  them  for $A_i\ , \ K $.\foot { Note that either $K$ or the
longitudinal part of $A_i$  will not be determined since one of them can be
always eliminated by a transformation of $v$ (2.3).}

The expressions for the   connection
and the curvature corresponding to
 (2.2) are generalisations of (2.14)--(2.19) [1,6,8] (in what follows we shall
return to the notation $x^\m$ for the coordinates in (2.2)). If $\hg_{\m \n } $
denotes the ``diagonal" part (2.1) of $\tg_{\m \n}$  then (see (2.17)--(2.19))
$$ \tg_{\m \n}=\hg_{\m \n}
 + 2 A_{(\m} l_{\n)}\ \ , \ \  \ \ \ \ A_\m \equiv (0,\ha K ,
A_i)  \ \ ,
  \eq{3.5} $$
$$ \ \  l_\m= \del_\m u \ \ , \ \  {\tilde  D_\m } l_\n =0 \ \ $$
 $$\tg^{\m \n}=\hg^{\m \n} - 2A^{(\m} l^{\n)} + A^2 l^\m l^\n  \ \
\ , \ \ \  A^\m=\hg^{\m \n} A_\n \ \ , $$
$$
{\tilde  \Gamma}^\l_{\mu \nu} = { \hat \Gamma}^\l_{\mu \nu } - ( \hg^{\l \r}-
A^\r l^\l)  F_{\r (\m} l_{\n)} + {\hat D}_{(\m} A_{\n ) } l^\l  \ \ ,
\ \ \ \ \ F_{\m \n} \equiv  2\del_{[\m} A_{\n ]} \ , \eq{3.6} $$
$$ {\tilde R}_{\m \n \r \s} ={\hat   R}_{\m \n \r \s} +  l_{[\m } {\hat D}_{\n
]
}F_{\r \s } +  l_{[\r } {\hat D}_{\s]} F_{\m \n } + l_{[\m } F_{\n] [\l }
F^\l_{ \ [\r }l_{ \s ] }  \ \ , \eq{3.7} $$
$$ {\tilde R}_{\m \n } = {\hat  R}_{\m \n } +l_{(\m } { D}^{\s } F_{\n ) \s }
+ \fourth l_\m l_\n F^{\r \s} F_{\r \s} \ \ .  \eq{3.8} $$
Explicitly,
 $$ {\tilde \Gamma}^i_{jk} = { \Gamma}^i_{jk}\ \ ,
 \ \ {\tilde \Gamma}^i_{ju}=\ha g^{im}( \dg_{jm} + F_{j m})   \ \ ,
\ \  {\tilde \Gamma}^i_{uu} = g^{im}({\dot A}_m - \ha \del_m K) \ \ , \ \
 {\tilde \Gamma}^v_{ij}=\ha \dg_{ij} - { D}_{(i} A_{j ) }
\ \ , $$ $$
{\tilde \Gamma}^v_{iu} = -\ha \del_i K + \ha A^m ( \dg_{im} + F_{im} )  \ \ ,
\ \  {\tilde \Gamma}^v_{uu}=  -\ha {\dot K}  + A^m ({\dot A}_m - \ha \del_m K)
\ \ , \ \  etc. $$
 In general, substituting (2.2) into (2.5)  we  find  again that the
dilaton  should  be linear in $v$ (2.20). The leading order terms in  the
$(iu)$
and $(uu)$ components of (2.5)    take the form (cf.(3.1),(3.2))
$$ \ha ( D^j \dg_{ij} - \del_i(g^{jk } \dg_{jk}))   + \ha D^j F_{ij} $$ $$
+ 2\del_i\dpp - (\dg_{ij} - F_{ij}) D^j \p    -  p F_{ik} A^k
  + p \del_i K +O(\a') =0 \ \ ,  \eq{3.9} $$ $$   - \ha
( g^{ij} \ddg_{ij}-\ha  g^{ij} g^{mn}\dg_{im}\dg_{nj})
- \ha  D^2 K  + \fourth F^{ij} F_{ij} + D^i{\dot A}_i $$ $$
+ 2 \ddp  -  2({\dot A}_i - \ha \del_i K ) (pA^i + 2D^i \p )  + p \dot K  +
O(\a')  =0 \ \ .  \eq{3.10} $$
Let us note also that the $(ij)$ component of (2.5)  and  eq.(2.6)  are
modified
as follows  (as compared to  (2.21),(2.22) and (2.25))
$$ \gb^G_{ij} +   D_{(i} W_{j )} + 2  D_{i} D_{j} \phi
  - p \dg_{ij}  + 2p D_{(i } A_{j)} =0 \ \
,  \ \ \eqno {(3.11)} $$
   $$ {1\over 3 } + {\bgb^{\phi}}{}' + \ha  p M^{ij} (\dg_{ij} -
2D_{(i}A_{j)})  + p^2( A^iA_i - K)
 - \ha p W_u - 2p \dpp
 =0  \ \ . \ \eqno {(3.12)} $$

3. Let us  now  consider the  special case when
$$\gij (u,x) = \gij (x) \ \ \ , \ \ \ \p
= \p (u) + \p' (x)  \ \ , $$ i.e.  $p=0$     and $\gij (x) \ , \ \p' (x)
$  represent a Weyl invariant theory with ${1\over 3 } + {\bgb^{\phi}}{}' =0$.
Then    we are  left with eqs. (3.9),(3.10), i.e.
 $$\ha D^j F_{ij}  +  F_{ij}  D^j \p + O(\a') = 0  \ \ , \eq{3.13} $$
$$ - \ha  D^2 K  + \fourth F^{ij} F_{ij} + D^i{\dot A}_i
+ 2 {\ddp}  -  2({\dot A}_i - \ha \del_i K )  D^i \p
+ O(\a')  =0 \ \ .  \eq{3.14} $$
The identity (2.30),(2.31)  can now be interpreted as   a  consequence of the
gauge invariance
$$A_i' = A_i - \del_ih \ \ \ , \ \ \ K' = K - 2 \dot h\ \ , \eq{3.15} $$  which
originates
from  the invariance under  redefinitions of $v$ (see (2.3),(2.4)).
To avoid (almost  all of) higher order corrections to (3.13),(3.14) let us
follow
refs.[3-8] and  further assume that  the transverse part of the metric is
flat,
$\gij = \d_{ij}$. If $p=0$ the condition that the  transverse theory should be
Weyl invariant  then implies that  $\p $ can be at most linear in $x^i$  and
for simplicity  we shall  set it equal to zero.  Then it is easy to see that
there are no $\a'$  corrections in (3.13) (the only terms that  may  contribute
to  $\b^G_{ui}$ could originate from the  structures  $D^sR$ which are linear
in
the curvature (3.7) but such higher order terms are actually absent in the
$\b^G_{\m \n}$ - function, cf.[6]). Possible higher order terms in (3.14) could
come from the terms  $D^sRD^rR$ in  $\b^G_{uu}$ which are quadratic in the
curvature and  therefore  will have the structure $\del^s F \del^r F$. As a
result, we are left with  the following system
 $$ \del^j F_{ij} =0 \ \ , \eq{3.16} $$
 $$ - \ha  \del^2 K  + \fourth F^{ij} F_{ij} + \del^i{\dot A}_i
+ 2 {\ddp}  + {1\over 8}  \a' \del_iF_{jk}
\del^iF^{jk} +  O({\a'}{}^s (\del^s F)^2) =0 \ . \eq{3.17} $$
The  exact solutions will be generated, for example,  by  solutions of (3.16)
for
which $F_{ij}$ is a polynomial  of finite degree in $x^i$ with $u$ - dependent
coefficients [6,7]. The simplest solution with $A_i=0$ was considered in [4,5].
A gauge equivalent (cf.(3.15)) solution with $F_{ij}=0$   corresponds to
 $$A_i=
a_{ij}(u) x^j\ \ , \ \ \ a_{ij}=a_{ji} \ \ , \eq{3.18} $$
where $a_{ij}$ is  an arbitrary symmetric  matrix (e.g.
$ a(u)\gd_{ij}$).  If  one starts with (2.1) with  $ \gij = \k (u) \gij (x)$
 then,  as it was found in [1], the Einstein equation is satisfied
if $\k  =  u^2 $. If $\gij (x)  = \d_{ij} \ , \p = 0$ this is an exact string
vacuum. Making a coordinate transformation to eliminate $\k (u)$ from the
transverse part of the
metric we get  (2.2) with $ A_i = -u^{-1}x_i \ , \ K= u^{-2} x^2 $ (cf.
(3.3),(3.4)), i.e. the equivalent solution of (3.16),(3.17).

   A less trivial example  of a solution of (3.16),(3.17)  is provided by
 $
F_{ij}= F_{ij} (u)$, i.e. by the plane - fronted wave metric [6-8]
 $$ A_i= -\ha F_{ij}(u) x^j \ \ , \eq{3.19} $$ $$ \ \ \    \del^2 K  -
\ha F^{ij} F_{ij}  -4 {\ddp}=0 \ \  ,  $$
$$ K =  k_{ij}(u) x^i x^j + k_0 \ \ , \ \ \  k^i_i = \fourth F^{ij} F_{ij} +2
{\ddp} \ \ .\eq{3.20}  $$
The equivalent metric  represented in the form (2.1)  (i.e. the equivalent
solution of (3.1),(3.2))  corresponds to a  special case of a flat  transverse
metric in (2.1),  namely  the $x^i$ - independent one, $\gij (u,x) = \gij (u)$.
In fact, the particular case of  the coordinate transformation (2.3),(2.4)
$$ x^i= L^i_j(u) {{x^j}{}'}\ \ , \  \ \ h =  s_{ij} x^i x^j \ \ , \eq{3.21}
$$
 where $L^i_j$ is expressed in terms of $F_{ij}$ relates (2.2) with
$\gij=\gd_{ij}$ and $A_i\ , \ K $ given by (3.19),(3.20) to (2.1) with  $ \gij
(u) = L^m_i L^n_j \gd_{mn}$.  Vice versa, if we start with (2.1) with  the
$x^i$ - independent metric  $\gij (u,x) = \gij (u)$ we can always make it
equal
to $\gd_{ij}$ by  a $u$-dependent  linear transformation  of the transverse
coordinates, $ x^i= (L^{-1})^i_j(u) y^j \ $ (the required
transformation is a particular case of (3.3),(3.4))
$$ ds^2 = -2dudv + \gij (u) dx^i dx^j \ \ , \ \  \gij (u) = L^m_i (u)L^n_j (u)
\gd_{mn}  \ \ . \eq{3.22} $$
  As a result, we get (2.2) with (we rename $y^i \rightarrow x^i \ $;
cf.(3.4))
$$ \gij=\gd_{ij}\ \ , \ \  A_i= f_{ij}(u) x^j \ \ , \ \  K= t_{ij} (u)  x^i x^j
\
\ , \ \  \eq{3.23} $$ $$
 f_{ij} = - \gd_{ik} ( {\dot L} L^{-1} ) ^k_j \ \ , \ \ t_{ij} =
 \gd_{mn} ( {\dot L} L^{-1} ) ^m_i( {\dot L} L^{-1} ) ^n_j \ \ , \ \ \
K= A_i A^i \ \ .    $$
This background is a solution of (3.16),(3.17) if (cf. (3.18)--(3.20))
$$ t^i_i = f_{[ij]}f^{[ij]} + {\dot f}^i_i  + 2 {\ddp} \ \ , \ \ \ \
\eq{3.24} $$
$$ F_{ij}=-2f_{[ij]} \ \ , \ \  a_{ij} = f_{(ij)} \ . $$
Eq.(3.24) is a second order equation for $L^i_j(u)$  equivalent to (3.2) (where
there are no higher order corrections if $\gij (u,x) = \gij (u)$).
Let us note that  as it is clear from (2.3),(2.4)   the solutions of
(3.9)--(3.12) with $\gij (u,x) = \gij (u)$ (considered in [7,8])  are gauge -
equivalent to the solutions with $\gij = \d_{ij}$ [6] discussed above.

  Being equivalent to   eqs. (2.21)--(2.25)   the system (3.9)--(3.12)
does not have   a non-trivial solution with flat transverse metric  in the
case when   the coefficient $p$ of the  $v$ term in the dilaton (2.20) is
non-vanishing.  In fact, if $p\not=0$  one  can   transform $v$ to absorb $\p
(u,x)$ in (2.20), i.e.
  to make the dilaton  equal to
$$ \p = \p_0 + pv \ \ \ .  \eq{3.25} $$
  If $\gij = \d_{ij}$ we  find  that  (3.9),(3.10),(3.11) and
(3.12)  reduce to (cf.(3.16),(3.17))\foot {Note that a $v$ - dependent dilaton
background breaks the gauge invariance (3.15).}
 $$    \del^j F_{ij}  -  2 p F_{ik}A^k   + 2p \del_i K  =0 \ \ ,  \eq{3.26} $$
$$  - \ha  \del^2 K  + \fourth F^{ij} F_{ij}
  - 2 p ({\dot A}_i - \ha \del_i K ) A^i  + p \dot K  +
O(\a')  =0 \ \ ,  \eq{3.27} $$
 $$   p \del_{(i } A_{j)} =0 \ \ \ ,  \ \ \eqno {(3.28)} $$
   $$ c + p^2( A^iA_i - K)  =0  \ \ , \ \eqno {(3.29)} $$
where we have already used (3.28)  to simplify (3.27),(3.29).
Since $p\not=0$ eqs.(3.26),(3.27) are  satisfied  as a consequence of
(3.28),(3.29). Eqs.(3.26)--(3.29) imply
that $ {\tilde R}_{\m \n } =0 $ and ${\tilde  \Gamma}^v_{\mu \nu} =0$ and hence
all the components of the curvature ${\tilde R}_{\m \n \l \r}$ vanish. Note
that
though (3.19) satisfies (3.28) the solution (3.19),(3.20) is non-trivial
since  in contrast to $K$ in (3.29) there in general $K\not= A_iA^i + k_0 $.

4. It is of interest to generalise the above discussion to the case of
non-vanishing antisymmetric tensor background. One of  motivations is that
WZW models or group spaces ``parallelised" by the
 antisymmetric tensor field strength [19] provide  simple explicit examples of
 conformally invariant backgrounds  which can be used  to represent the
transverse theory. One would like also  to find solutions describing
interpolation in $u$ between different conformal points.
If the sigma model action contains  also the antisymmetric tensor  $B_{\m \n}$
coupling
$$I= {1\over { 4 \pi \a' }} \int d^2 z \sqrt {\g}\ [ \ (G_{\m \n} + B_{\m \n})
(
\g^{ab} + i \ep^{ab}) \del_a x^\m \del_b x^\n \  +  \  R^{(2)} \gp \ ] \  \ ,
$$ then the leading terms in the Weyl anomaly coefficients  are given by
[19,15](cf.(2.5),(2.6)) $$ {{\bar \gb}^G}_{\mu \nu } =\a'( R_{\mu \nu} -
\fourth
H_{\m \l \r}  H_\n^{\  \l \r} + 2  D_{\mu} D_{\nu} \phi ) +  O(\a'^2)=0 \ \ ,
\ \ \ H_{\l \m \n } \equiv 3\del_{[\l}B_{\m \n]} \ \ , \eq{3.30} $$
$$  {{\bar \gb}^B}_{\mu \nu } = -\ha \a' D^\l H_{\l \m \n }  +
\a' \del_\l \p H^\l_{\ \m \n} + O(\a'^2)=0
\ \ ,  \ \   \eq{3.31} $$
$$ \bgb^{\phi} = c - \ha \a' D^2 \p   + \ga'(\del_\mu \phi)^2 - {1\over 24 }
\a'
H_{\l \m \n}^2  +  O(\a'^2)=0  \ \ . \ \eqno {(3.32)} $$
Let us assume that in addition to the metric (2.1) or (2.2) we are given a $v$
- independent $\B_{\m \n} $ background
 $$ \B_{ij} = B_{ij}(u,x)\ \ , \ \ \  \B_{iu} = B_i (u,x)\ \ , \eq{3.33} $$
$$\ \ \
\ \H_{ijk}= H_{ijk} \equiv  3\del_{[i}B_{jk]}\ \ , \ \  \H_{uij}= H_{ij} \ \ ,
\
\ \hij \equiv \dB_{ij} + 2\del_{[i }  B_{j ]}   \ \ , \eq{3.34} $$
or, in `covariant' notation (cf.(3.5),(3.6))
$$ \B_{\m \n} = B_{\m \n} + 2 B_{[\m } l_{\n ] } \ \ , \ \ \ \H_{\l \m \n} =
H_{\l \m \n} + 3l_{[\l} H_{\m \n ]} \ \ ,  \eq{3.35} $$
where $B_{\m \n}\ , \ B_\m \ , \ H_{\l \m \n }$ and $H_{\m \n}$ have only
transverse components  being non-vanishing. The remaining gauge symmetry
$$ B_{ij}'= B_{ij} + 2 \del_{ [ i}  \l_{j ] } \ \ , \ \  B_i'= B_i +
\del_i \l_u - \del_u \l_i \ \ , \ \  \l_u=\l_u (u,x)\ \ , \ \ \l_i=\l_i (u,x)\
\
,\eq{3.36} $$ allows  one to  absorb  $B_i$  into $B_{ij}$ ($B_i$ plays the
role
similar to that of $A_i$ in (2.2)). Eqs.(3.30),(3.32)  written in components
take the form ($\a'=1$)
$$ {{\bar \gb}^G}_{ij} = R_{ij} - \fourth H_{i m n }
H_j^{ mn }  +... =0\ \ , \eq{3.37} $$
$$ {{\bar \gb}^G}_{ui} =R_{ui} - \fourth H_{mn}
H_i^{ mn}  + ... =0\ \ , \eq{3.38} $$
$$ {{\bar \gb}^G}_{uu} =R_{uu} - \fourth H_{mn}
H^{ mn}  + ... =0\ \ , \eq{3.39} $$
$$ {{\bar \gb}^B}_{ij } = -\ha  D^m H_{mij }  +
 \del_m \p H^m_{\ ij} - p (\dB_{ij} + 2\del_{[i }  B_{j ]})   + ... =0 \ \ ,
\eq{3.40} $$ $$ {{\bar \gb}^B}_{iu } = -\ha  D^m H_{mi }  +
 \del_m \p H^m_{\ i}   +   \ha  {\tilde  \Gamma}^m_{nu} H^m_{\ ni} +  ... =0 \
\
, \eq{3.41} $$
where ${\tilde  \Gamma}^m_{n u}= \ha g^{mk}\dg_{kn} - \ha F^m_{\ n} \ $  (see
(3.6),(2.17),(2.14)) and we have assumed that $\p$ is given by (2.20)
(cf.(2.21)--(2.25)).  Like eq.(2.21)  ${{\bar \gb}^B}_{ij } =0$  (eq.(3.40))
can
be interpreted (for $p\not= 0$) as the  renormalisation group equation for the
coupling $B_{ij}(u,x)$  of the transverse theory.

Let us  consider the case when
$$ \gij (u,x) = \k (u) g_{ij}(x) \ \ , \  \ B_{ij}
(u,x) = q(u) b_{ij} (x) \ \ ,  \eq{3.42} $$
where $g_{ij}$ and $b_{ij}$ correspond to a group space  and are normalised in
such a way that the curvature of the generalised connection with torsion
$${\bar \Gamma}^i_{jk} \equiv  {\Gamma}^i_{jk}  - \ha H^i_{\ jk} = \ha g^{im} (
\del_j \r_{mk} + \del_k \r_{jm} - \del_m \r_{jk}) \ , \ \ \ \ \r_{ij} \equiv
g_{ij} + b_{ij}  \eq{3.43} $$
 vanishes. Then  equations (3.11),(3.37) and (3.40)  take
the form
$$ p {\dot \k} g_{ij}  - 2p D_{(i }  A_{j )}  =  a(1- \k^{-2} q^2 )
g_{ij}  + ... \ \ , \ \ \ \ \ a \equiv  R/N \ \ , \eq{3.44} $$
$$ p  {\dot q} b_{ij} + 2p\del_{[i }  B_{j ]} =0  \ \ ,
$$ where we have assumed  that  the dilaton is homogeneous ($\p = \p (u)$)
and used that the group space torsion is covariantly constant.  If $p\not=0$
one  should  put $A_i=0\ , \ B_i=0$  obtaining
$$ p {\dot \k} =  a(1- \k^{-2} q^2 ) + ...  \ \ ,  \ \  \ \ \  p  {\dot q}=0 \
\
. \eq{3.45} $$
There exists a renormalisation scheme in which higher order terms in (3.45)
are also proportional to $  1- \k^{-2} q^2$.
The conformal fixed point corresponds to $\k= \pm q =const$ [19].
This point is unstable: if one starts with $\k= \pm q $ at $u=0$  one
finds $\k (u)  \rightarrow u$ at large $u$.
If we   take the transverse theory  to be at the conformal point
   for all $u$ and set
$p=0$ ($\p = \p (u)$) then  eqs.(3.37) and (3.40) (and (3.32)) are satisfied
automatically so that  we  are left with eqs.(3.38),(3.39),(3.41) for
 $A_i\ , \ B_i$  and  $K$.\foot { It is interesting to note  a similarity
in  the structure of eqs. (3.38) and  (3.41) for $A_i$ and $B_i$ which
suggests  to look for solutions with  $A_i=B_i$.}
 Since in general there are no invariant vector and
scalar functions on the group space there  seems to be  no non-trivial
solutions.

One can  obtain a  simple  set of solutions by generalising those with
flat transverse space   to the presence of the ``trivial"
antisymmetric tensor background represented by $B_i$ [3,5,6]. The case of
``homogeneous" antisymmetric tensor $B_{ij} = B_{ij} (u) $ is equivalent to
the case of  $B_{ij} = 0$    because of gauge invariance (3.36) (one can
absorb  $B_{ij} (u)$ into  $B_{i}$ by the redefinition $B_i \rightarrow
B_i - \ha \dB_{ij } x^j $).
If
$$\gij = \gd_{ij}\ , \ B_{ij}=0\ , \ \p = \p (u)\ , \ p=0$$
 we get  from
(3.38),(3.41),(3.39) the following system of equations for $A_i , \  B_i $ and
$K$  (cf. (3.16),(3.17))
$$ \del^j F_{ij} =0 \ \ , \eq{3.46} $$
$$ \del^j H_{ij} =0 \ \ ,  \ \ \hij =  2\del_{[i }  B_{j ]} \ \ ,
\eq{3.47} $$
 $$ - \ha  \del^2 K  + \fourth F^{ij} F_{ij}  - \fourth H^{ij} H_{ij}
+ \del^i{\dot A}_i
+ 2 {\ddp}  +   O({\a'}{}^s (\del^s F)^2 , \ {\a'}{}^s(\del^s H)^2 ) =0 \ .
\eq{3.48} $$
 Note that  as in (3.16),(3.46) there are no higher order corrections in
eq.(3.47). The simplest solution of (3.47) is provided by $ B_i= - \ha H_{ij}
(u) x^j $ [3,5]. This solution is equivalent to the one with  $B_{ij} =
B_{ij} (u) \ , \  B_{i} = 0$.
%%%%%%%%%%%%%%%%%%%%%%%%%%%%%%%%%%%%%%%%%%%%%%%%%%%%%%%%%
 \newsec{ Another   representation of  generic
solution and connection with $2d$ quantum gravity model}

A relation between the model (2.1) and $2d$ quantum gravity coupled to a
`transverse' sigma model was already pointed out  in [9]. Below we shall
further
clarify this
 relation using a slightly different version of the basic solution discussed
in Sec.2.

1.  In sec.2 we  were solving the Weyl invariance conditions (2.5),(2.6) using
the metric (2.1). As we noted, (2.1) is the most general ansatz for a null
metric if it is understood that  some particular choice of coordinates $x^i$
and $v$ have already being made.  One may try instead to look for solutions  in
terms of the metric (2.2) assuming that  the freedom  to redefine  $v$  was
used
to fix  the form of the dilaton (2.20)   as in  (3.25). In fact, the  metric -
dilaton background $\gij (u,x)\ , \ \p= pv + \p (u,x)  $
 is equivalent to $\gij (u,x)\ , \ A_i =p^{-1} \del_i \p (u,x) \ , \ K= 2
p^{-1} \dpp (u,x) \
, \ \p= p v  $.   Let us  consider  an $inequivalent$ solution of
(3.9)--(3.12)  for which $ \ A_i =0$ but $K\not =0$, i.e.
$$ds^2 =  -2dudv +  \ggij (u,x) dx^i dx^j
 + K(u,x) du^2  \ \ \ , \ \ \ \p = p v \  \ .  \ \eq{4.1} $$
 When $A_i=0$  eqs.(3.5)--(3.8) simplify as follows [3,5]:
 $$ \tg_{\m \n}=\hg_{\m \n}
 + Kl_{\m} l_{\n}\ \ , \ \
\tg^{\m \n}=\hg^{\m \n} - Kl^{\m} l^{\n} \ \ , \
 \ A_\m = \ha K l_\m \ \ , \ \ \  \eq{4.2} $$ $$
{\tilde  \Gamma}^\l_{\mu \nu} = { \hat \Gamma}^\l_{\mu \nu } -  \ha \hg^{\l
\r}  \del_{\r } K l_{\m}  l_{\n} + {\del}_{(\m} K l_{\n ) } l^\l  \ \ , \ \
F_{\m \n} = \del_{[\m} K l_{\n ]} \ \ ,  \ \eq{4.3} $$
 $$
{\tilde R}_{\m \n \r \s} ={\hat   R}_{\m \n \r \s} +  2l_{[\m } {\hat D}_{\n ]
}{\hat D}_{[\r}K l_{\s ]} \ \ , \ \  {\tilde R}_{\m \n } =
{\hat  R}_{\m \n }   - \ha {\hat D}^2 K l_\m l_\n  \ \ . \eq{4.4} $$
Then   (3.11) and
(3.12)  take the following form (cf. (2.21), (2.22), (2.25))
 $$ \gb^G_{ij} +   D_{(i} W_{j )}
- p \dg_{ij}  =0 \ \ ,  \ \ \eqno {(4.5)} $$
   $$ {1\over 3 } + {\bgb^{\phi}}{}' + \ha  p M^{ij} \dg_{ij}  - \ha p W_u
 - p^2 K =0  \ \ . \ \eqno {(4.6)} $$
Since $p\not=0$ the remaining equations (3.9),(3.10)  (cf.(2.23),(2.24))
$$ \gb^G_{iu} + \ha \del_i W_u + \ha \dot W_i - \dg_{ij}W^j +
p \del_i K =0 \ \ ,  \eq{4.7} $$
$$ \gb^G_{uu} + \dot W_u  - \ha W^i \del_i K
  + p \dot K =0 \ \   \eq{4.8} $$
should again be satisfied as a consequence of (4.5),(4.6)
(note that all $K$ - dependence in (4.7),(4.8)
is shown explicitly). Substituting $\dg_{ij}$ from (4.5) into (4.6) we find the
following
expression for $K$ in terms of functions of $\gij$ only
$$ p^2 K = {1\over 3 } + {\bgb^{\phi}}{}' +
\ha  M^{ij} (\gb^G_{ij} +   D_{(i} W_{j )})     - \ha p W_u \ \
 . \eq{4.9} $$
 Since (4.5) is a first order equation for $\gij (u,x)$ and $K$ is
explicitly given by (4.9) we conclude that  the system (4.5)--(4.9) always
has a solution for  generic initial conditions.

  It is
interesting to note that $K$  given by  (4.9) has  a natural  interpretation as
the basic  ``central charge" Weyl anomaly coefficient of the transverse theory
(a
linear combination of ${\bgb^{\phi}}{}'$ and $\bgb^g_{ij}$) which is
changing  with `time' $u$.\foot{ Eq. (4.8) giving the expression for $\dot K$
may be related to the $c$ - theorem [20]. The embedding  of the RG flow of a
non-conformal $N$ - dimensional  ``transverse" theory into the Weyl invariance
conditions of the $N+2$ - dimensional theory [9] may help to  clarify
the meaning of the $c$ - theorem  in the
sigma model context (cf.[16]).}  For example, in the leading order
approximation
(4.7) and (4.9) take the form
 $$   p \dg_{ij} = R_{ij} + O(
\a') \ \ , \ \eq{4.10} $$
$$ p^2 K = c + \fourth R  +  O( \a') \ \ . \eq{4.11} $$
It is easy to check that eqs.(4.7),(4.8), namely,
 $$ \ha ( D^j \dg_{ij} - \del_i(g^{jk }
\dg_{jk}))
  + p \del_i K +O(\a') =0 \ \ ,  \eq{4.12} $$
 $$   - \ha
( g^{ij} \ddg_{ij}-\ha  g^{ij} g^{mn}\dg_{im}\dg_{nj})
- \ha  D^2 K   + p \dot K  +
O(\a')  =0 \ \ ,  \eq{4.13} $$
are indeed satisfied identically on (4.10),(4.11).

2. The  sigma model corresponding to (4.1) ($\a'=1$)
 $$ I= {1\over { 4 \pi  }} \int d^2 z \sqrt {\g}\
 [\  G_{\mu \nu
}(x) \del_a x^{\mu} \del^a x^{\nu}  +   R^{(2)} \gp (x)\ ] \  $$
$$= {1\over { 4 \pi  }} \int d^2 z \sqrt {\g} [\
- 2 \del_a v \del^a u  +  \gij (u,x) \del_a x^{i} \del^a x^{j}
+ K(u,x) \del_a u \del^a u   + pv R^{(2)} \ ] \ \ , \eq{4.14} $$
 may be interpreted  as   a  ``quantum action"  (represented in the conformal
gauge) of the scalar-tensor $2d$ gravity  theory  coupled to the transverse
sigma
model. In fact,  consider the following  classical action
$$ I_0 = {1\over { 4 \pi  }} \int d^2 z \sqrt {\hgg} [\ p v \hR^{(2)} +
\gij (x) \del_a x^{i} \del^a x^{j} \ ]\ \ , \eq{4.15} $$
 where $v(z)$ is an extra scalar field coupled to $2d$ gravity (see e.g.
[10--13]). In the conformal gauge
 $$ \hgg_{ab} = \e{-2u/p} \g_{ab} \ \ , \eq{4.16} $$
 (4.15)  takes the form
$$ I_0 = {1\over { 4 \pi  }} \int d^2 z \sqrt {\g} [\
- 2 \del_a v \del^a u  +  \gij (x) \del_a x^{i} \del^a x^{j}
 + p v R^{(2)}\ ] \ \ . \eq{4.17} $$
Since
$u(z)$ is  proportional to the conformal factor of the $2d$ metric one expects
that at the quantum level  $\gij $ (which in general depends on a cutoff)
should
start running with $u$ according to the RG equation (4.10). Also,  the
conformal anomaly term ($\sim (\del u )^2 $) should appear. The total theory
should be conformal invariant with respect to the background metric $\g_{ab}$
since the $2d$ metric itself  is an integration variable [21--23]. This is
precisely the result we  have got in (4.14) with $K$  playing indeed the role
of
the ``central charge" coefficient of the transverse ($N$ - dimensional) sigma
model!

 In principle, one could expect   the
quantum action to contain  also another  anomaly  structure  $ \p(u,x)
R^{(2)}$. However, as  we have seen, the condition of  conformal invariance of
the theory (4.14)  is   satisfied without need to introduce  such term.   If
  to represent the quantum analog of (4.15) we have used instead
of (4.1) the solution of Sec.2 (cf.[9])
then  the conformal invariant quantum action   would contain the dilaton
term $ \p (u,x)  R^{(2)}$  instead of the ``anomaly" term $  K(u,x) \del_a u
\del^a u $,
  $$I= {1\over { 4 \pi  }} \int d^2 z \sqrt {\g} \ [\
- 2 \del_a v \del^a u  +  \gij (u,x) \del_a x^{i} \del^a x^{j}
  + pv R^{(2)}+ \p (u,x)  R^{(2)} \ ] \ \ . \eq{4.18} $$
The important difference  between  the theory  (4.15)  and  the standard $2d$
gravity coupled to a sigma model (where one expects both  the anomaly term  $
K(u,x) \del_a u \del^a u $ and $ \p (u,x)  R^{(2)}$ to appear in the quantum
action [23]) is due to the presence of the  extra  scalar field $v$.   Though
it   could seem that introducing an extra field
we would  make the theory  $N+2$-dimensional,  it, in fact,   remains
effectively  $N+1$ - dimensional as in the absence of $v$ since the couplings
do
not depend on $v$ (the Killing symmetry is preserved by renormalisation).

 It is straightforward to generalise the above discussion to the case when
the tachyon coupling $T$ is included into the sigma model action, i.e. when
there is a potential term in (4.14).  The Weyl invariance condition
corresponding
to $T$ has the standard form [24,17,25] (cf.(2.5)--(2.8))
$${\bar \beta}^T=-\g T + ( \a' \del^\mu \phi + \ha W^\mu ) \del_\mu T
-2T +  b(T) $$
$$ = - \ha \a' D^2 T + \a' \del^\mu \phi \del_\mu T  -2T + O(\a'^3) + b(T)
=0 \ \ , \eq{4.18} $$
where $\g$ is the same operator as in (2.8) and $b(T)$ denote
``non-perturbative"
corrections which are of higher order in $T$ (similar terms are present in
(2.5),(2.6)). If the metric $G_{\m \n}$ is null (2.1) and the dilaton is linear
in $v$ (2.20) then for $v$-independent  tachyon $T=T(u,x)$ eq.(4.18) takes the
form
similar to (2.21),(2.25),(3.40), i.e. becomes RG-type equation which
 is first order in the $u$-derivative
$$ p{\dot T} = {\bar \b}^T_x \ \ \ . \eq{4.19} $$
${\bar \b}^T_x$ denotes the Weyl anomaly coefficient of the ``transverse"
theory with coupling $T(u,x)$ and $u$ playing the role of the RG ``time"
(${\bar \b}^T_x$ does not depend on $v$ and contains only derivatives over
$x^i$).
To provide the simplest example of a solution of (4.18),(4.19) let us
drop the dependence on $x^i$  and ignore for a moment
 ``non-perturbative" terms. Then
$$ p {\dot T}= 2 T \ \ \ , \ \ \ \  T = T_0 \e{ 2 u/ p } \ \ . \eq{4.20} $$
Equivalent solution in the context of $2d$ gravity model was discussed in
the last two papers in [12]. It is interesting to note that this solution
is actually the exact one, i.e. it solves  (4.18) with all higher order terms
included.
In fact, it is easy (as compared to the case of the Liouville theory)
 to see that there are no non-perturbative divergences
in the model
$$I = {1\over { 4 \pi  }} \int d^2 z \sqrt {\g}\  [\
- 2 \del_a v \del^a u  +
pv R^{(2)} + T (u) \ ] \ \ . \eq{4.21} $$
$v$ plays the role of a Lagrange multiplier (for flat $\g_{ab}$ background)
  so that $u$ is effectively non-propagating. As a result, there are
essentially
no quantum corrections in the theory (a similar observation was made in [26]).
Then the condition of conformal invariance is equivalent to
 the classical conformal invariance relation (4.20).\foot { One may wonder
how to reconcile this conclusion with the expected
presence of $O(T^2)$  and $O(\del T \del T )$ terms
in ${\bar \b}^T$ and ${\bar \b}^G$.  As discussed in [25], the derivation of
such
terms (which correspond to analogous terms in the effective action) presumes
analytic continuation in momenta and is not, strictly speaking, valid in the
case when $T$ depends on just one variable. The question of non-perturbative
terms in the $\b$-functions should be studied separately in each particular
theory (e.g. Liouville, sin-Gordon or (4.21)). }

In general, it appears that the $2d$ gravity model with an extra scalar field
  is better defined
and simpler than pure $2d$ gravity (which does not have a  non-trivial
tree level action).  What we have  demonstrated above is that  one can
describe
the coupling of   quantum  $2d$  scalar -  gravity  system to a
non-conformal  theory in terms of a  conformally invariant   sigma model  in
$N+2$  dimensions. Similar representations
 for models  describing coupling of pure $2d$ quantum
gravity to  a  non-conformal matter
in terms of $N+1$ dimensional conformal theories
are not  explicitly known. Moreover, in an
attempt to find such a representation one runs into  the problem of
ambiguities
in choosing proper initial conditions since the corresponding evolution
equations are $second$ order in  `time' (conformal factor).  It is
remarkable that by introducing   one extra dimension but at the same time
imposing the null Killing symmetry it is possible to  interpret the  conformal
invariance conditions on a higher dimensional theory  as the standard  (first
order) RG equations for the matter theory.
%%%%%%%%%%%%%%%%%%%%%%%%%%%%%%%%%%%%%%%%%%%%%%%%%%%%%%%

\newsec{Concluding remarks}
In this paper we have studied solutions of the string effective equations for
the backgrounds with covariantly constant Killing vector.  We have generalised
the  previous discussions [3--8] to the case when the transverse theory is
non-conformal and
the dilaton contains the term linear in light cone coordinate $v$. The
resulting equations can be interpreted as the RG equations for the couplings of
the transverse theory [9]. We have proved the existence of the solutions by
making use of the  general covariance identities  for   the Weyl anomaly
coefficients [14]. We have also clarified the question of gauge equivalence of
different backgrounds and  reproduced  the solutions of refs.[3--8]  in a
systematic way.

We have suggested the connection between the solutions (conformal invariant
$2+N$ - dimensional sigma models) and the  $2d$ scalar quantum gravity coupled
to a non-conformal `transverse' $N$ - dimensional  sigma model. The conformal
factor of the $2d$ metric  is identified not with  time but with the light cone
coordinate $u$.  The difference as compared to ref.[9] is that we  have used
the
solution for which the  analog of the conformal anomaly term appears  in the
sigma model action.  It would be interesting to clarify further the  connection
between   the corresponding ``propagating" conformal theories and
$2d$ quantum gravity models.

\bigskip

I  am  grateful to G. Gibbons  and G. Horowitz
for useful discussions and, in particular, to G. Gibbons
for   emphasising  to me the  relevance of ref.[1]  and for drawing my
attention
to refs.[8].  I would like to thank the International School for Advanced
Studies (SISSA), Trieste and the Aspen Center for Physics for
 hospitality  while parts of this work were done.
I would like also to acknowledge the support of Trinity College, Cambridge.

%%%%%%%%%%%%%%%%%%%%%%%%%%%%%%%%%%%%%%%%%%%%%%%%%%%%%%%%%%%%%%

\vfill\eject

\centerline{\bf References}
\bigskip

\item {[1]} H.W. Brinkmann, Math. Ann. 94(1925)119 .

\item {[2]} D.Kramer, H. Stephani, E.Herlt and M. MacCallum, {\it  Exact
solutions of

Einstein's Field Equations} (Cambridge U.P., 1980).

\item {[3]} R. Guven, Phys. Lett. B191(1987)275 .

\item {[4]} D. Amati and C. Klimcik, \pl B219(1989)443 .

\item {[5]} G. Horowitz and A.R. Steif, Phys. Rev. Lett. 64(1990)260 ;

Phys. Rev. D42(1990)1950 .

\item {[6]}G. Horowitz, in: {\it Proceedings
of  Strings '90},

  College Station, Texas, March 1990 (World Scientific,1991).

\item {[7]}R.E. Rudd, \np B352(1991)489 .

\item {[8]} C.Duval, G.W. Gibbons and P.A. Horv\'athy, \pr D43(1991)3907 ;

 C.Duval, G.W. Gibbons, P.A. Horv\'athy and M.J. Perry, unpublished (1991) .

\item {[9]} A.A. Tseytlin, \pl B288(1992)279 .

\item{[10]}
C. Teitelboim, \pl B126(1983)41 ;

R. Jackiw, in: {\it Quantum Theory of Gravity}, ed. S.Christensen (Adam

Hilger, Bristol 1984) ;

A.H. Chamseddine, Phys. Lett. B256(1991)2930; \np B368(1992)98 ;

T. Banks and M. O'Loughlin, Nucl. Phys. B362(1991)649 .

\item{[11]}C.G. Callan, S.B. Giddings, J.A. Harvey and
A. Strominger,

\pr D45(1992)1005 .

\item{[12]} J. Russo and A.A. Tseytlin, \np B382(1992)259 ;

H. Verlinde, preprint PUPT-1303 ;

A. Strominger, preprint UCSBTH-92-18 ;

T.T. Burwick and A.H. Chamseddine, preprint ZU-TH-4/92 ;

S.P. deAlwis, preprint COLO-HEP-280(1992) ;

A. Bilal and C. Callan, preprint PUPT-1320 ;

S.B. Giddings and A. Strominger, preprint UCSBTH-92-28 .

\item {[13]}  B. de Wit, M.T. Grisaru, E. Rabinovici and H. Nicolai,

\pl B286(1992)78 .

\item {[14]}  G. Curci and G. Paffuti, \np B268(1987)399 .

\item {[15]} C.G. Callan, D. Friedan, E. Martinec and M.J. Perry, \np

B262(1985)593 .

\item {[16]} A.A. Tseytlin, Int. J. Mod. Phys. A4(1989)1257 .

\item {[17]} A.A. Tseytlin, \pl B178(1986)34 ; \np B294(1987)383 .

\item {[18]} G.M. Shore, \np B286(1987)349 ;

H. Osborn, \np B294(1987)595 .

\item {[19]} E. Witten, Commun. Math. Phys. 92(1984)455 ;

E. Braaten, T.L. Curtright and C.K. Zachos, \np B260(1985)630 ;

S. Mukhi, \pl B162(1985)345 .

\item {[20]} A.B. Zamolodchikov, JETP Lett. 43(1986)349 .

\item {[21]} F. David, Mod. Phys. Lett. A3(1988)1651 ;

     J. Distler and H. Kawai, Nucl.Phys. B321(1989)509 ;

     S.R. Das, S. Naik and S.R. Wadia, Mod. Phys. Lett. A4(1989)1033 ;

     J. Polchinski, Nucl. Phys. B324 (1989)123 .
\item {[22]} T. Banks and J. Lykken, \np B331(1990)173 .
\item {[23]} A.A. Tseytlin, Int. J. Mod. Phys. A5(1990)1833 .

\item {[24]} C. Callan and Z. Gan, \np B277(1986)647 .
\item {[25]} A.A. Tseytlin, \pl B241(1990)233 ; \pl B264(1991)311 .
\item {[26]} J. Russo, L. Susskind and L. Thorlacius, preprint SU-ITP-92-24 .

 \vfill
\eject
\bye